\documentclass[twocolumn,showpacs,preprintnumbers,amsmath,amssymb]{revtex4}

\usepackage{xspace}
\usepackage{graphicx}
\usepackage{dcolumn}
\usepackage{bm}

\def\beq{\begin{equation}}
\def\eeq{\end{equation}}

\def\bea{\begin{eqnarray}}
\def\eea{\end{eqnarray}}

\newcommand{\roughly}[1]%
    {{\mathrel{\raise.3ex\hbox{$#1$\kern-.75em\lower1ex\hbox{$\sim$}}}}}
\newcommand{\lsim}{\mathrel{\roughly<}}
\newcommand{\gsim}{\mathrel{\roughly>}}

\newcommand{\tr}{\mathop{\rm tr}}

\newcommand{\Ga}{\ensuremath{\Gamma}}

\newcommand{\De}{\ensuremath{\Delta}}
\newcommand{\ep}{\ensuremath{\epsilon}}

\newcommand{\ka}{\ensuremath{\kappa}}
\newcommand{\la}{\ensuremath{\lambda}}

\newcommand{\si}{\ensuremath{\sigma}}

\newcommand{\Om}{\ensuremath{\Omega}}

\newcommand{\sfrac}[2]{{\textstyle\frac{#1}{#2}}}

\newcommand{\avg}[1]{\ensuremath{\langle{#1}\rangle}}

\newcommand{\GeV}{\ensuremath{\mathrm{~GeV}}}
\newcommand{\TeV}{\ensuremath{\mathrm{~TeV}}}

\newcommand{\Eq}[1]{Eq.~(\ref{#1})}

\newcommand{\Ref}[1]{Ref.~\cite{#1}}

\begin{document}


\title{Leptonic Indirect Detection Signals\\
from Strongly Interacting Asymmetric Dark Matter}

\author{Yi Cai}
\email{yicai@ucdavis.edu}
 \author{Markus A. Luty}%
 \email{luty@physics.ucdavis.edu}
\affiliation{%
Physics Department, University of California Davis\\
Davis, California 95616}%

\author{David E. Kaplan}%
\email{dkaplan@pha.jhu.edu}
\affiliation{Physics Department, Johns Hopkins University\\
Baltimore, Maryland 21218}


\begin{abstract}
Particles with TeV mass and strong self-interactions generically have
the right annihilation cross section to explain
an observed excess of cosmic electrons and positrons
if the end-product of the annihilation is charged leptons.
We present an explicit model of strongly-coupled TeV-scale dark matter
whose relic abundance related to the matter-antimatter asymmetry
of the observed universe.
The $B - L$ asymmetry of the standard model is transfered to the
dark sector by an operator carrying standard model lepton number.
Lepton number violation naturally induces dark matter
particle-antiparticle oscillations at late times,
allowing dark matter-antimatter annihilations today.
The dark matter annihilates into lighter strongly-interacting
particles in the dark sector that decay to leptons via the
transfer operator.
The strong dynamics in the dark sector is at the weak scale
due to supersymmetry breaking.
The correct dark matter abundance can be obtained for a wide
range of couplings with all masses in the dark sector at the
TeV scale.
\end{abstract}

\pacs{12.60.Jv, 95.35.+d}

\maketitle

Recent astrophysical data on high-energy cosmic electrons and positrons
from {\it e.g.\/}~PAMELA \cite{PAMELA}, FERMI \cite{FERMI},
and HESS \cite{HESS} show excesses over expected backgrounds that may
be evidence for dark matter annihilation in our galaxy
\cite{indirectDMreview}.
Although there are possible astrophysical explanations for these
observations \cite{astro} it is important to explore possible particle
physics interpretations of these effects, since these have implications
for experiments both in astrophysics and colliders.
The absence of a high energy falloff in the FERMI spectrum
requires the dark matter mass to be a TeV or larger.
If we assume standard distribution of dark matter in our galaxy, the observed rate requires an annihilation cross section $\avg{\si_{\rm ann} v} \sim \mbox{nb}$.
This is just what one expects for a heavy composite particle in a theory with strong interactions at the TeV scale.
If the strong scale is $M$, heavy composite particles have mass $M$
and annihilation cross section at low velocities given by
\beq
\label{strongann}
\si_{\rm ann} \sim \frac{4\pi}{M^2 v},
\eeq
where the $1/v$ arises from $s$-wave enhancement.  An alternative mechanism to explain the large event rate
is infrared enhancement due to attractive interactions (Sommerfeld enhancement) \cite{Sommerfeld}.
\Eq{strongann} assumes that there is no such enhancement.  It is striking that the correct mass and cross section are obtained
in this way, and the TeV mass scale suggests that the strong interaction
scale is tied to the scale of electroweak symmetry breaking.

We will
show that this possibility can be naturally realized in a
compelling framework for dark matter and electroweak physics.
The dark matter relic abundance is due to a
particle-antiparticle asymmetry in the dark matter sector
\cite{oldADM,ADM}.
We will consider supersymmetric theories
in which the standard model $B - L$ asymmetry
is transferred to the dark sector via 
superpotential couplings such as
\beq
\label{Wtrans}
W_{\rm trans} \sim X^2 L H_u.
\eeq
This higher-dimension interaction
is naturally in equilibrium at high temperatures
and drops out of equilibrium as the universe cools,
freezing in a separate asymmetry
in the dark matter and visible sectors.
We follow \Ref{ADM} in calling this scenario
``asymmetric dark matter.'' 

In the early universe dark matter annihilation
goes out of equilibrium due to the absence of antiparticles.
The annihilation cross section can therefore
be much larger than in conventional WIMP
models where the dark matter is a thermal relic.
It would appear that in this scenario there is no dark antimatter
today to annihilate and give a direct detection signal.
However, neutrino oscillations strongly suggest that $B - L$
is broken by small Majorana neutrino masses.
If one assumes the same for the dark matter,
small dark matter Majorana masses will give rise to
dark matter-antimatter oscillations at late times.
These can erase the dark matter asymmetry and allow
for annihilation today in the galactic halo.

If the interactions that transfer the asymmetry
freeze out at a temperature $T_f$ larger than the dark matter
mass $M$, the dark matter asymmetry
is of order the baryon asymmetry, and we obtain
\beq
\Om_{\rm DM} \sim \Om_B \frac{M}{m_p}
\quad
(T_f > M),
\eeq
which gives the correct dark matter abundance for $M \sim 10\GeV$ \cite{ADM}.  If $T_f < M$ then the dark matter asymmetry is diluted by a Boltzmann factor, and we obtain
\beq
\Om_{\rm DM} \sim \Om_B \frac{M}{m_p}
e^{-M/T_f}
\quad
(T_f < M).
\eeq
A more precise calculation (see below)
shows that $T_f \simeq M/7.6$ gives the correct relic abundance for
$M \simeq \mbox{2 TeV}$.
Decoupling at this scale requires interactions that are much
weaker than perturbative interactions with dimensionless
order-1 couplings.
In our scenario this arises from a naturally small
dimensionful coupling.
Although we do not predict the relic density, the correct
relic density is obtained for a wide range of values
for the dimensionful coupling.
%

The final ingredient required for this scenario is strong dynamics at the TeV scale in the dark matter sector.
The strong sector must have a global $U(1)_X$ symmetry whose
conserved charge $X$ distinguishes between dark particles and 
antiparticles.
The interaction \Eq{Wtrans} induces an $X$ asymmetry from the
standard model $B - L$ asymmetry.
The lightest particle charged with $X \ne 0$ (``baryon'') 
is stable, and is naturally the dark matter particle. 
The strong sector also generally contains lighter particles (``pions'')
with $X = 0$, so dark matter-antimatter pairs can efficiently
annihilate into these states.
The transfer operator in
\Eq{Wtrans} explicitly breaks both $U(1)_X$ and $U(1)_{B-L}$, but
leaves a linear combination intact.
In the presence of $\De L = 1$ breaking of lepton number,
a $Z_2$ subgroup of remains exact, which ensures that
the dark matter is absolutely stable.
The lightest of the neutral dark states naturally
decay through the same operator that transferred the asymmetry.
This gives decays to leptons, precisely the signal we are
interested in.

The TeV scale is the scale of electroweak symmetry breaking,
suggesting that the strong dynamics in the dark sector is connected
with electroweak physics.
We give a mechanism for this in the context of supersymmetry (SUSY),
which gives a compelling framework for physics at the TeV scale. 
We assume that the dark sector is a strongly-coupled SUSY conformal theory
above the TeV scale.
In many SUSY breaking scenarios, SUSY breaking masses are at the TeV scale in both the visible and the dark sector.
The SUSY breaking masses in the dark sector trigger conformal symmetry
breaking, and can give rise to strongly-coupled non-SUSY dynamics at
the TeV scale.
We present an explicit model below.

We now discuss the points above in more detail,
starting with the freeze-in of the dark matter asymmetry.
This occurs when the interaction 
\Eq{Wtrans} freezes out at a temperature
$T_f < M$ where the dark matter is non-relativistic.
The equilibrium asymmetry induced 
in the MSSM is then given by
\beq
\frac{X}{B - L} = \frac{88}{79}
\left( \frac{M}{2\pi T} \right)^{3/2} e^{-M/T}
\eeq
where the left-hand-side is a ratio of charge densities and we define the 
$X$ charge of the $X$ particle to be $+1$.
The exponential suppression is due to the asymmetry being transfered
from the dark sector into the visible sector.
The relic dark matter density is then
\beq
\frac{\Om_{\rm DM}}{\Om_B} = 
\frac{M}{m_p} \frac{X}{B}
\simeq 0.23\, \frac{M}{m_p} 
\left( \frac{M}{T_f} \right)^{3/2} e^{-M/T_f},
\eeq
where we 
assume that the asymmetry transfer operator freezes out
above the electroweak phase transition,
and use $B / (B - L) \simeq 0.31$
to obtain the numerical estimate.
For $M = \mbox{2 TeV}$ we
obtain the correct dark matter density
for $T_f \simeq M/7.6$.

We now discuss dark matter oscillations.
Neutrino oscillations give a strong hint that $B - L$
is broken by small Majorana neutrino masses, and is therefore
not an exact symmetry. 
It is very natural for the dark matter particle
to have small Majorana masses (we assume the dark matter is a fermion).
In a supersymmetric theory, the mass terms arise from
\beq
\label{Wmass}
W_{\rm mass} = M \tilde{X} X
+ \sfrac 12 \mu_X X^2 + \sfrac 12 \tilde{\mu}_X \tilde{X}^2.
\eeq
The Majorana mass terms break $B - L$ and
give rise to dark matter-antimatter oscillations with frequency
$|\mu_X + \tilde{\mu}_X| \ll M$.
These can restore equal abundance of dark matter and antimatter today.
In order for this mechanism to be effective, the oscillations
must not restore equal number of particles and antiparticles
before a temperature where annihilation with a nb cross section becomes
ineffective at restoring equilibrium between dark matter and antimatter.
The oscillation time must also be
shorter than the lifetime of the universe.
This gives
$10^{-33}~\mbox{eV} \ll |\mu_X + \tilde{\mu}_X| \lsim 10^{-11}~\mbox{eV}$
for $M \sim \mbox{TeV}$.
These scales are much smaller than neutrino mass scales.
However Majorana neutrino masses are $\De L = 2$, and these
will not generate $X$ Majorana masses of these are $\De L = 1$.


We now discuss the dynamics of the strongly-coupled SUSY conformal
dark sector.  
The theory must have at least one global
$U(1)$ symmetry to identify with $U(1)_X$.
Also, in the presence of SUSY breaking the theory must have a
stable vacuum (no runaway) that preserves $B - L$.
This requirement is very nontrivial, as we will see.
Conceptually, the simplest such theory is one where the only scale
is the SUSY breaking scale.
This occurs if the strong sector contains no weak couplings or
large $N$ factors, and has no light states such as
Nambu-Goldstone bosons, chiral fermions, or massless $U(1)$ gauge bosons.
We are unable to either find an example or prove that this is impossible.
We instead present a model with weak couplings that stabilize the VEVs.
This serves as an existence proof for strong dark sectors.
Given our lack of knowledge of strong SUSY conformal theories,
it is reasonable to assume that there are many other possiblities.

The strong sector of our model consists of an $SU(2)$ gauge theory with an $SU(2)$ global symmetry and fields $\Phi_1, \ldots, \Phi_4$ transforming as $(\frac 12, \frac 12)$.
The superpotential can be written in terms of
gauge-invariant ``meson'' fields $(\Phi_i \Phi_j)$, which is a two-by-two matrix in $SU(2)$ flavor, and where the strong indices are contracted
with $\ep = i\si_2$.
The theory has a superpotential
\beq
\label{Wstrong}
W_{\rm strong} \sim \mathop{\rm tr}\left[ (\Phi_1 \Phi_2) \ep (\Phi_3 \Phi_4) \ep) \right]
+ \mathop{\rm tr}\left[ (\Phi_2 \Phi_3) \ep (\Phi_4 \Phi_1) \ep) \right]
\eeq
invariant under a global $SU(2) \times U(1)^3 \times U(1)_R \times Z_4$,
where $Z_4$ cyclically permutes $\Phi_1, \ldots, \Phi_4$.
All of these symmetries will be broken by perturbative interactions and SUSY breaking, as described below, save a $Z_2$ which will keep dark matter from decaying.
The quartic superpotential \Eq{Wstrong} is believed to be
exactly marginal \cite{LeighStrassler},
so there is plausibly a fixed point where it is strong.
We will assume the theory is at such a fixed point.
Strong conformal dynamics is known to suppress
all soft SUSY breaking except those that can be written as SUSY breaking components of a spurion gauge superfield of a global symmetry \cite{SUSYbreakCFT}.
The dominant SUSY breaking in this sector
is therefore a scalar mass proportional to an anomaly-free generator
in the strong sector.
We assume SUSY breaking masses $m_i^2$ for the fields $\Phi_i$ with $m_1^2 + \cdots + m_4^2 = 0$ and
$m_{1,2}^2 > 0$, $m_{3,4}^2 < 0$.

In the SUSY limit, this theory has flat directions where $(\Phi_3 \Phi_4) \ne 0$.  Along these flat directions, all other mesons get massive.  The SUSY breaking masses produce a runaway direction along this flat direction that must be stabilized by additional interactions.  We add superpotential couplings of the form
\beq
\label{Wweak}
\De W \sim  
\tr\left[(\Phi_3\Phi_3)^2\right]
+ \tr\left[(\Phi_4\Phi_4)^2\right]
+\left( \tr\left[(\Phi_3 \Phi_4)\right]\right)^2
+ \cdots 
\eeq
These break the global symmetries of the strong sector, and are therefore not exactly marginal: they have dimensionless coefficients with logarithmic renormalization.
We assume the theory contains the most general quartic superpotential that preserves $U(1)_{B-L}$, where $\Phi_{1,2}$ have $B - L = \pm \frac 14$ and $\Phi_{3,4}$ have $B - L = 0$.
Preserving this global $B - L$ symmetry appears somewhat arbitrary, but can arise as an accidental symmetry if $B - L$ is a gauge symmetry broken at high energy to a discrete ($Z_3$ or larger) subgroup.  
The stabilization of the VEVs depends on the K\"ahler potential
for the light fields, which depends on the strong dynamics and is
not known.
We have checked that that superpotential terms \Eq{Wweak} are sufficient to stabilize $\avg{\Phi_{3,4}}$ with maximal rank using K\"ahler potential
$K \propto (\sum_i \mathop{\rm tr} \Phi_i^\dagger
\Phi^{\vphantom\dagger}_i)^{4/3}$,
which at least has the correct scale dimension.
This breaks the strong gauge symmetry completely.
The scale of strong dynamics from the VEV is of order
\beq
M \sim \frac{4\pi m}{\la},
\eeq
where $m$ is the size of the SUSY breaking soft mass and $\la$ is the size of dimensionless couplings in \Eq{Wweak}.  The factor of $4\pi$ comes from requiring that the theory is strongly coupled at the scale $M$
in the sense that tree and loop graphs with strong couplings
are of the same order \cite{NDA}.

The VEVs break $U(1)_R$, so this theory has a massless $R$ axion
in the SUSY limit.
However, $U(1)_R$ is broken by SUSY breaking terms.
$U(1)_R$ is explicitly broken by unsuppressed $B$-like terms
(a soft-breaking scalar quartic) proportional to a $U(1)$ charge that is preserved by the strong superpotential \Eq{Wstrong}.
This gives a quartic potential from the weak superpotential (which is not invariant under the global symmetries), and therefore an $R$ axion mass
$m_R^2 \sim \la M m / 4\pi \sim m^2$
assuming $B \sim m$.

The strong superpotential \Eq{Wstrong} gives a mass of order $M$ to all states carrying $B - L$.  In the SUSY limit, the states carrying $B - L$ are the ``heavy-heavy'' composite states 
$(\Phi_1\Phi_1)$ and $(\Phi_2 \Phi_2)$,
and the ``heavy-light'' states $(\Phi_i\Phi_3)$, $(\Phi_i\Phi_4)$, $i,j = 1,2$.  Which of these states is the lightest depends on dynamics and the breaking of the global symmetries and SUSY.  It is possible that there are several kinematically stable states carrying $B - L$ that can contribute comparably to dark matter.  The details do not affect the phenomenology at the level discussed here, and we will assume for definiteness that the dark matter is a fermion with quantum numbers $(\Phi_1\Phi_3)$.

We now discuss the freeze-out of the
interactions that transfer the $B - L$ asymmetry
to the dark sector at $T_f \simeq M/7.6$.
For definiteness we will assume a transfer operator
\Eq{Wtrans} with $X = \tr(\Phi_1 \Phi_3)$.
The UV completion of the transfer operator requires additional
massive fields.
It is particularly natural for their mass to arise from the
same mechanism that generates the $\mu$ term in the MSSM,
in which case they have a mass in the TeV range.
We add new gauge singlet particles $A$ and $N$ with couplings
\begin{eqnarray}
W_{\rm trans} &=& \sqrt{\ka}\, A \tr(\Phi_1 \Phi_3) + M_A A \bar{A}
+ h \bar{A}^2 N 
\nonumber\\
\label{WtransTeV}
&& \qquad + M_N N \bar{N} + \ep \bar{N} L_\tau H_u.
\end{eqnarray}
Near the fixed point, the first term is a relevant interaction, and
$\ka$ has units of mass.
In order for this term to be a perturbation to the conformal
dynamics, we require $\ka \lsim M \sim \mbox{TeV}$.
We assume $h \sim 1$, and a flavor suppression
$\ep \sim \sqrt{y_\tau} \sim 10^{-1}$.

We assume that $M_N, M_A \sim \mbox{TeV}$
so the rate-limiting processes are naturally those that
are not exponentially suppressed at $T = T_f$.
The coupling $\ka$ naturally provides the small parameter
we need to suppress the processes that are not exponentially
suppressed.
It is therefore natural for the limiting process to be
$X L \leftrightarrow \bar{X} \bar{H}_u$.
The rate per $X$ particle is given by
\begin{eqnarray}
\Ga_X(X L \leftrightarrow \bar{X} \bar{H}_u)
&\sim& n_L \si(X L \to \bar{X} \bar{H}_u) v
\nonumber\\
\label{kapparate}
&\sim& \frac{T^3}{\pi^2}
\times \frac{1}{16\pi} 
\left( \frac{\ep \ka M}{M_N M_A^2} \right)^2
\end{eqnarray}
For $M_N, M_A \sim 10\TeV$ this decouples at
the correct temperature for $\ka \sim 5\GeV$.
Of the remaining processes, the most important is
$A \leftrightarrow \mbox{CFT}$,
which has a rate per $X$ particle
\beq
\Ga_X(A \leftrightarrow \mbox{CFT})
\sim 
\left( \frac{M_A}{2\pi M} \right)^{3/2}
\frac{\ka}{16\pi}
e^{-(M_A - M)/T}.
\eeq
This also falls out of equilibrium at $T \sim T_f$
for the parameters described above.
To illustrate that this model can give the right relic
abundance for a wide range of parameters, we take
$\ka \sim 10^{-4}\GeV$, the smallest value allowed
by decays of the dark sector into leptons (see below).
Then $A \leftrightarrow \mbox{CFT}$
gives the right relic abundance for $M_N \sim 10\TeV$,
$M_A \sim 7\TeV$,
while $X L \leftrightarrow \bar{X} \bar{H}_u$
is negligible.

\begin{figure}
\includegraphics{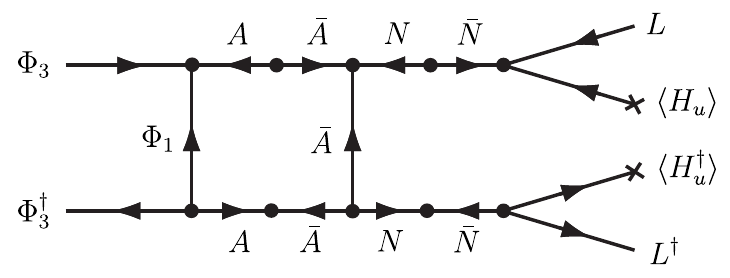}
\caption{\label{fig:epsart} Feynman diagram giving rise to decay
of light composite mesons in the dark sector.}
\end{figure}

We now discuss the annihilation of dark matter into the visible sector.
The 1-loop diagram in Fig.~1 generates an effective 
K\"ahler interaction that allows light mesons to decay:
\beq
\Ga(\Pi \to \tilde{\nu}\tilde\nu,
\tilde{\Pi} \to \nu\tilde\nu)
\sim \frac{1}{16\pi}
\frac{\ka^2\ep^4}{(16\pi^2)^2} 
\frac{v_u^4 M m^3}{M_A^5 M_N^4}.
\eeq
The decay $\Pi \to \nu\nu$ is suppressed by the neutrino mass.
The leading sneutrino decay naturally involves charged leptons
({\it e.g.\/}\ $\tilde\nu \to \ell \tilde{W}$),
giving the signal we are interested in.
This decay length is shorter than the typical distance
traveled by TeV positrons in our galaxy ($\sim\mbox{kpc}$) for
$\ka \gsim 10^{-4}\GeV$.
Since the leptons can carry only a fraction of the total
energy, we need $M \gsim 2\TeV$.

We now briefly discuss experimental signals of this model.
In addition to the charged lepton signal, there is a 
possible high-energy neutrino signal from $\tilde\Pi$
decays.
This may be observable \cite{neutrinoDM},
but the rates depend sensitively on the dark matter
profile near the core of the galaxy.
The Majorana mass $\mu_X$ breaks $R$ parity, so 
the lightest supersymmetric particle (LSP)
is unstable, but the lifetime induced from
$\mu_X$ is at least $10^{12}$ times longer than
the lifetime of the universe, and there are no bounds
on such decays.
The LSP may have a shorter lifetime if
$R$ parity breaking is more strongly coupled
to the visible sector.
The collider signatures are those of supersymmetry,
possibly with observable LSP decays.
It appears impossible to produce the dark sector particles
at colliders because of their large mass and weak coupling
to the standard model.
For the same reason, we expect no observable direct detection
signal in this model.

There are many directions for future work.
On the model-building side, it is interesting to
consider other possibilities for the dark sector,
{\it e.g.\/}~5D models
``dual'' to strong conformal sectors.
On the phenomenological side, it is
interesting to work out details of the
observed spectra for charged leptons and
neutrinos in these models.
While this work was in progress,
\Ref{CZ} appeared, which investigates many of the same ideas.

{\it Acknowledgements---}%
We thank K. Zurek for collaboration in earlier stages of this
investigation.
We thank S. Chang for discussions.
M.A.L. thanks the
Aspen Center for Physics, where part of this work was
carried out.
Y.C. and M.A.L. are supported by DOE contract
DE-FG03-91ER40674, and
D.E.K. was supported by NSF grant NSF-PHY-0401513.

\end{document}